\documentclass[aps,prl,reprint,nofootinbib,superscriptaddress,longbibliography,floatfix]{revtex4-2}

\usepackage{xcolor}
\usepackage{amsmath,amssymb,bm,mathrsfs}
\usepackage{graphicx}

\definecolor{mPink}{HTML}{C46D82}
\definecolor{mGreen}{HTML}{5E8B7E}
\definecolor{mBlue}{HTML}{4A90A4}
 
\definecolor{mPurple}{HTML}{7A6993}  
\usepackage[colorlinks=true,linkcolor=mPink,citecolor=mGreen,urlcolor=mBlue,filecolor=mPurple]{hyperref}

\newcommand{\ii}{\mathrm{i}}
\newcommand{\dd}{\mathrm{d}}
\newcommand{\wtom}{\widetilde{\omega}}

\begin{document}

\title{Boundary spectroscopy of hidden near-horizon reflection}

\author{Yiling Chen}
\email{yilingchen@jxnu.edu.cn}
\affiliation{School of Physics, Jiangxi Normal University, Nanchang 330022, China}
\author{Jiayue Yang}
\email{j43yang@uwaterloo.ca}
\affiliation{Department of Applied Mathematics, University of Waterloo, 200 University Ave W, Waterloo, ON N2L 3G1, Canada}
\affiliation{Institute for Quantum Computing, University of Waterloo, 200 University Ave W, Waterloo, ON N2L 3G1, Canada}
\affiliation{Perimeter Institute for Theoretical Physics, 31 Caroline St N, Waterloo, ON N2L 2Y5, Canada}
\affiliation{Department of Physics and Astronomy, University of Waterloo, 200 University Ave W, Waterloo, ON N2L 3G1, Canada}
\author{Zhen Zhong}
\email{zhen.zhong@uniroma1.it}
\affiliation{Dipartimento di Fisica, Sapienza Università di Roma, Piazzale Aldo Moro 5, 00185, Roma, Italy}
\affiliation{INFN, Sezione di Roma, Piazzale Aldo Moro 2, 00185, Roma, Italy}
\author{Robert B. Mann}
\email{rbmann@uwaterloo.ca}
\affiliation{Department of Physics and Astronomy, University of Waterloo, 200 University Ave W, Waterloo, ON N2L 3G1, Canada}
\affiliation{Perimeter Institute for Theoretical Physics, 31 Caroline St N, Waterloo, ON N2L 2Y5, Canada}
\affiliation{Institute for Quantum Computing, University of Waterloo, 200 University Ave W, Waterloo, ON N2L 3G1, Canada}
\affiliation{Department of Applied Mathematics, University of Waterloo, 200 University Ave W, Waterloo, ON N2L 3G1, Canada}
\author{Jie Jiang}
\email{jiejiang@bnu.edu.cn}
\affiliation{Faculty of Arts and Sciences, Beijing Normal University, Zhuhai 519087, China}
\author{Ming Zhang}
\email[Corresponding author, ]{mingchang@outlook.com}
\affiliation{School of Physics, Jiangxi Normal University, Nanchang 330022, China}

\begin{abstract}
A classical black hole horizon absorbs completely, and any deviation would be hidden in the bulk. For a scalar field in a rotating Ba\~nados-Teitelboim-Zanelli exterior, we derive an exact boundary response of a corotating passive wall near the horizon. An Airy filter gives its spectrum, and positive energy excludes growing modes. On any frequency band the response identifies a constant wall impedance and the wall position, which enter orthogonally with weights determined by the spectral function. The boundary thus sees a reflecting wall, whereas a perfect absorber is exponentially hidden.
\end{abstract}

\maketitle

\textcolor{mBlue}{\textit{Introduction---}}
The purely ingoing condition at the future horizon provides the standard classical description of black hole absorption \cite{Son:2002sd,Berti:2009kk}. Motivated by proposals for quantum horizon structure and by exotic compact objects, we parameterize departures by a local passive wall close to the horizon \cite{Cardoso:2016rao,Abedi:2016hgu,Nakano:2017fvh,Conklin:2017lwb,Wang:2018gin,Cardoso:2019rvt,Oshita:2019sat,Wang:2019rcf,Oshita:2020dox}. For a black hole in anti-de Sitter (AdS) spacetime, holography connects bulk propagation and infrared boundary conditions to the response of a boundary conformal field theory (CFT) \cite{Son:2002sd,Birmingham:2001pj,Kovtun:2005ev}. For a weak source $\mathcal J$ coupled to a boundary operator $\mathcal O$, the response in frequency space is $\delta\langle\mathcal O\rangle=G_R\mathcal J$, with $G_R$ the retarded Green function in the source convention specified below. Experiments on critical spin chains now resolve conformal spectra and boundary effects \cite{Sun:2026geo}, which makes spectroscopy operational. Can boundary spectroscopy diagnose reflection hidden deep in the bulk, and individually resolve the loss and the position of the reflector?

Black hole quasinormal modes (QNMs) characterize oscillation frequencies and damping rates \cite{Nollert:1999ji,Kokkotas:1999bd,Berti:2009kk}, their residues determine how strongly a source excites them, nonresonant contributions such as branch cuts and redshift terms influence the signal \cite{Rosato:2026moe,Mark:2017dnq}, and in holography the contact terms require a renormalization convention \cite{Skenderis:2002wp}. Complete response constructions already relate resonance denominators to multiple returns \cite{Mark:2017dnq,Testa:2018bzd,Correia:2018apm}, and absorption spectra show cavity filtering, with the logarithmic amplitude loss controlling the linewidths \cite{Macedo:2018yoi}. A local dissipative wall of this kind yields a semiclassical loss relation in a reflecting Schwarzschild-AdS cavity \cite{Xie:2026yjg}, and a viscous membrane provides a similar impedance boundary for dark compact objects in flat space \cite{Maggio:2020jml}. Studies of semiopen systems mapping pole migration and the scattering response \cite{Solidoro:2024yxi,Wu:2025wbp}, nonnormal black hole spectra are highly sensitive to perturbations of the potential or operator \cite{Jaramillo:2020tuu,Cheung:2021bol,Arean:2023ejh,Cownden:2023dam}, and under analyticity assumptions the thermal product formula determines a two-sided thermal correlator from its QNMs up to normalization \cite{Dodelson:2023vrw}. Here we seek an exact response at finite absorption that separates a hidden reflection amplitude from exterior dispersion, and we ask what such a response can identify.

The Ba\~nados-Teitelboim-Zanelli (BTZ) geometry \cite{Banados:1992wn,Banados:1992gq} offers an exact map between horizon modes and boundary sources and responses, together with the thermal CFT$_2$ correlator \cite{Keski-Vakkuri:1998gmz,Son:2002sd,Birmingham:2001hc,Cardoso:2001hn,Birmingham:2001pj}. Brick-wall mode counting \cite{tHooft:1984kcu}, scalar wave falloff \cite{Chan:1996yk}, thermal decay and recurrences \cite{Solodukhin:2005qy,Maldacena:2001kr,Barbon:2003aq,Festuccia:2006sa}, and echoes in AdS black holes \cite{Saraswat:2019npa}, in horizonless AdS geometries \cite{Bena:2019azk,Dimitrov:2021csq,Chen:2025jbf}, and in Kerr/CFT \cite{Dey:2020wzm} motivate examining horizon structure from the boundary. Dirichlet brick-wall studies have determined normal modes, level structure, emergent thermality, and boundary Green functions, including rotation \cite{Kay:2011np,Iizuka:2013kma,Banerjee:2024dpl,Das:2022evy,Krishnan:2023jqn,Das:2023xjr,Banerjee:2024ivh,Burman:2023kko,Das:2024fwg,Gayari:2026jwn}, and stretched-horizon degrees of freedom can make such a wall absorbing with residual reflection \cite{Terashima:2025tct,Terashima:2026fix}. A membrane condition on the BTZ stretched horizon fails to capture the quasinormal modes \cite{deBoer:2014xja}. The distinction here, also from dissipative cavities in Schwarzschild-AdS \cite{Xie:2026yjg}, is the exact finite-radius response in rotating BTZ, a positive-energy delay identity applicable to all masses above the Breitenlohner-Freedman (BF) bound, and the identification of the wall.

In this Letter, we regard the exterior (spacetime geometry and the scalar mass) and the boundary conventions (the source normalization and the contact terms) as known, and derive the exact boundary response of a scalar field outside a local passive wall corotating with the horizon. Its spectral function is the spectrum of a perfectly absorbing wall multiplied by a positive Airy filter, with inverted resonances in the superradiant band, and a positive energy excludes growing modes. The complex value of the response at regular real frequencies determines the induced horizon return amplitude and, given the wall position, the wall impedance. Without the position, the response on any band still fixes both position and impedance, as long as the impedance is constant. We give the exact conditioning of this inverse problem, in which loss and position decouple, loss information is zero at corotation, and a perfectly absorbing wall is located only at near-horizon order.

\textcolor{mBlue}{\textit{Exact boundary response and reflection map---}}
We use units $c=\hbar=k_B=1$. The rotating BTZ metric is
\begin{align}
\dd s^2={}&-N^2\dd t^2+N^{-2}\dd r^2
+r^2(\dd\phi+N^\phi\dd t)^2,\nonumber\\
N^2={}&\frac{(r^2-r_+^2)(r^2-r_-^2)}
{L^2r^2},
\quad
N^\phi=-\frac{r_+r_-}{Lr^2},
\label{eq:metric}
\end{align}
where $L$ is the AdS radius, $N$ and $N^\phi$ are the lapse and shift functions, and $r_+>r_-\geq0$ are the horizon radii. The horizon angular velocity and surface gravity are $\Omega_H=r_-/(Lr_+)$ and $\kappa=(r_+^2-r_-^2)/(L^2r_+)$, and the left and right temperatures of the dual CFT are $T_{L,R}=(r_+\mp r_-)/(2\pi L^2)$, with Hawking temperature $T=2T_LT_R/(T_L+T_R)=\kappa/(2\pi)$. For a scalar field of mass $\mu$, the left and right conformal weights are
$
h_L=h_R=h=\left(1+\sqrt{1+\mu^2L^2}\right)/2,
$
and the dual operator has dimension $\Delta=2h$ \cite{Birmingham:2001pj}. For scalar masses in the BF window $-1<\mu^2L^2<0$ \cite{Breitenlohner:1982jf,Breitenlohner:1982bm}, we adopt standard quantization.

We consider a scalar field $\varphi=e^{-\ii\omega t+\ii m\phi}R(r)$ satisfying $(\Box-\mu^2)\varphi=0$, with azimuthal number $m\in\mathbb Z$ and radial profile $R(r)$. We then introduce the radial coordinate $z=(r^2-r_+^2)/(r^2-r_-^2)$, the boundary momentum $k=m/L$, and the dimensionless frequencies
\begin{equation}
\alpha_L=\frac{\omega-k}{4\pi T_L},
\quad
\alpha_R=\frac{\omega+k}{4\pi T_R},
\quad
\alpha=\alpha_L+\alpha_R=\frac{\tilde\omega}{\kappa},
\label{eq:frequencies}
\end{equation}
where $\tilde\omega=\omega-m\Omega_H$ is the frequency measured in the corotating frame. The exact solutions $u_-$ and $u_+$ are respectively ingoing and outgoing at the event horizon and are normalized to unit horizon amplitude, with $u_\mp\sim z^{\mp\ii \alpha/2}=e^{\mp\ii\wtom x}$ in the horizon phase coordinate $x=(\ln z)/(2\kappa)$ \cite{Keski-Vakkuri:1998gmz}. Their source and response coefficients $a_\sigma$ and $b_\sigma$ are defined by the boundary expansion $u_\sigma\sim a_\sigma(1-z)^{1-h}+b_\sigma(1-z)^h$ and take the form as
\begin{equation}
\begin{aligned}
a_\sigma&=\frac{\Gamma(1+\ii\sigma \alpha)\Gamma(2h-1)}
{\Gamma(h+\ii\sigma\alpha_L)\Gamma(h+\ii\sigma\alpha_R)},\\
b_\sigma&=\frac{\Gamma(1+\ii\sigma \alpha)\Gamma(1-2h)}
{\Gamma(1-h+\ii\sigma\alpha_L)\Gamma(1-h+\ii\sigma\alpha_R)}
\label{eq:coeffs}
\end{aligned}
\end{equation}
for generic noninteger $2h$, with $\sigma=-1,+1$ labeling the ingoing and outgoing channels, respectively. Because $1-z\sim(r_+^2-r_-^2)/r^2$, the two powers correspond to the $r^{\Delta-2}$ source and $r^{-\Delta}$ response falloffs. The determinant of the transfer matrix linking horizon with boundary in \eqref{eq:coeffs} obeys the exact bilinear identity
\begin{equation}
a_-b_+-a_+b_-=-\frac{\ii \alpha}{2h-1}.
\label{eq:det}
\end{equation}
At corotation, $\alpha=0$, the horizon solutions merge and the determinant vanishes, and we use the boundary basis $v_\beta$ below. For integer $2h$, logarithmic terms require finite-part renormalization \cite{Skenderis:2002wp}, and we restrict to noninteger $2h>1$.

We place the wall at a fixed radius $r_w>r_+$ and define its location by $z_w=z(r_w)$ and $x_w=\ln z_w/(2\kappa)<0$. The wall corotates with the horizon, as a stretched horizon or membrane does \cite{Price:1986yy}, so that its time generator is $\chi=\partial_t+\Omega_H\partial_\phi$, and it has a constant dimensionless impedance $q$. The exact wall condition is $(\partial_x-q\chi)\varphi|_{r_w}=0$ with $q\geq0$. The spectral and weak-loss formulas below use the branch $q\leq1$, with $q>0$ at finite loss. The  boundary operator evaluated at $r_w$ is $B_q=\partial_x+\ii q\wtom$. The wall replaces the horizon condition on $r\geq r_w$, with the dissipative derivative taken in the wall frame.

Let $v_{1-h}$ and $v_h$ individually be the radial solutions with unit leading source and response coefficient, we have
\begin{equation*}
v_\beta=z^{-\ii\alpha/2}(1-z)^\beta\,{}_2F_1(\beta-\ii\alpha_L,\beta-\ii\alpha_R;2\beta;1-z),
\end{equation*}
where $\beta=1-h,h$. In the holographic description, the coefficient of the slower falloff $(1-z)^{1-h}$ is the source of $\mathcal O$ and that of $(1-z)^h$ its response \cite{Son:2002sd,Skenderis:2002wp,Iqbal:2008by}. The solution obeying the wall condition is $U=v_{1-h}+G_qv_h$, so $B_qU=0$ gives the response per unit source, the exact reduced response $G_q$. The retarded Green function adds a normalization and a contact term determined by by the action (End Matter),
\begin{equation}
\begin{aligned}
G_R(\omega,k)&=-\mathcal N_hG_q(\omega,k)+P(\omega,k),\\
G_q(\omega,k)&=-\frac{B_qv_{1-h}}{B_qv_h}
=\frac{b_-+\zeta_{\rm loc}b_+}{a_-+\zeta_{\rm loc}a_+},
\end{aligned}
\label{eq:exactG}
\end{equation}
where $G_R(\omega,k)=\int dt\,d\phi\,e^{\ii\omega t-\ii m\phi}G_R(t,\phi)$ and $G_R(t,\phi)=-\ii\theta(t)\langle[\mathcal O(t,\phi),\mathcal O(0,0)]\rangle$ is the retarded commutator for the source coupling $H_{\rm src}=\int d\phi\,\mathcal J\mathcal O$, $\mathcal N_h\propto(4h-2)(r_+^2-r_-^2)^{2h-1}$ is positive, and $P$ is a local contact term. For a contact term that is real on the real frequency axis, $\rho_{\rm phys}\equiv-2\operatorname{Im}G_R=\mathcal N_h\rho_q$, where $\rho_q\equiv2\operatorname{Im}G_q$ is the reduced spectral function. We display reduced responses in units of $\mathcal N_h$ and omit $P$. The second form of $G_q$ is the same solution in the horizon basis, $U\propto u_-+\zeta_{\rm loc}u_+$, with $\zeta_{\rm loc}=-B_qu_-/B_qu_+$ the outgoing amplitude induced at the horizon per unit ingoing amplitude ($\zeta_{\rm loc}=0$ for the ingoing horizon). The wall law is the junction condition of an outgoing channel of impedance $Z_c$ connected at $r_w$, with $q=Z_c/r_+$, and the commutator is that of the closed system of exterior and channel (End Matter).

To reveal the cavity structure, we define local wall amplitudes $C_{\rm in}$ and $C_{\rm out}$ by $R(r_w)=C_{\rm in}+C_{\rm out}$ and $\partial_xR(r_w)=-\ii\wtom(C_{\rm in}-C_{\rm out})$. The wall condition gives $C_{\rm out}=K_wC_{\rm in}$, with amplitude reflectivity $K_w=(1-q)/(1+q)$. The range $q\leq1$ corresponds to $0\leq K_w\leq1$, reflection without a sign reversal, while passivity holds for every $q\geq0$ and $q\to\infty$ is the Dirichlet brick wall. Both $q$ and $K_w$ refer to the $(\partial_x,\chi)$ normalization. Channels normalized in the tortoise coordinate give the high frequency reflectivity $(r_w-qr_+)/(r_w+qr_+)$, which differs from $K_w$ by $O(z_w)$. The corresponding radial channels $w_\pm$, normalized to unit amplitude at the wall, have $w_\pm(r_w)=1$ and $\partial_xw_\pm(r_w)=\pm\ii\wtom$, with boundary coefficients $a_{w\pm},b_{w\pm}$. The solution is $U\propto w_-+K_ww_+$, so $G_q=(b_{w-}+K_wb_{w+})/(a_{w-}+K_wa_{w+})$. So $K_w$ is to the wall channels what $\zeta_{\rm loc}$ is to $u_\pm$. With no AdS source, the exterior return per unit outgoing wall amplitude is
$\mathcal R_w=-a_{w+}/a_{w-}=(\ii\wtom v_h-\partial_xv_h)/(\ii\wtom v_h+\partial_xv_h)$.
Defining the response $G_{\mathrm{abs},w}=b_{w-}/a_{w-}$ of the finite-wall absorber (the $q=1$, $K_w=0$ wall at $r_w$) and the channel response ratio $V_w=b_{w+}/b_{w-}$, $G_q$ in terms of the wall channel can be written as
\begin{equation}
G_q=G_{\mathrm{abs},w}
\frac{1+K_wV_w}{1-K_w\mathcal R_w}.
\label{eq:factorized}
\end{equation}
All propagation from the wall to the asymptotic AdS region and back is encoded in $\mathcal R_w$. At regular frequencies the cavity pole condition is
\begin{equation}
K_w\mathcal R_w(\omega_\nu,k)=1,
\label{eq:poleeq}
\end{equation}
where $\omega_\nu$ is a complex resonance frequency and $\nu\in\mathbb Z$ indexes the discrete poles. For $K_w=0$ the poles are those of the finite-wall absorber while at $K_w=1$ they become real Neumann cavity modes. Zeros satisfy $B_qv_{1-h}=0$ and residues and cancellations are analyzed in the End Matter.

The same response also yields an inverse map. At frequencies where the horizon basis is nonsingular, $\zeta_{\rm loc}=(b_--G_qa_-)/(G_qa_+-b_+)$. Alternatively, the local relation $q=\partial_xU/(-\ii\wtom U)|_{r_w}$ recovers the impedance when $\wtom U(r_w)\neq0$, in the spirit of the radial flow of boundary conditions \cite{Faulkner:2010jy}. Both formulas require the exterior and the boundary conventions, and the impedance formula also requires the wall position. They recover the phase of the induced return, whereas the total loop phase also includes the exterior reflection. Applied across frequencies, this calibrated inversion checks whether the reconstructed impedance is real and constant. Note that the response loses sensitivity to $q$ at corotation.

For a wall close to the horizon, $u_\pm=e^{\pm\ii\wtom x_w}[w_\pm+O(z_w)]$, so that at fixed frequencies away from degeneracies $\zeta_{\rm loc}=K_we^{-2\ii\wtom x_w}+O(z_w)$. This constant horizon ratio is a useful approximation on the real axis, but for $K_w>0$ the response built from it has upper half-plane poles near $m\Omega_H+\ii\kappa n$ (for $m\neq0$ at every sufficiently large $n$, see End Matter). The thermal response $G_0=b_-/a_-$ is recovered from $G_{\mathrm{abs},w}$ only as the wall approaches the horizon \cite{deBoer:2014xja}.

\textcolor{mBlue}{\textit{Exact spectral function and the superradiant band---}}
The hidden reflector first manifests itself as an interference filter on the absorbing boundary spectrum. On the real frequency axis the $v_\beta$ are real, so $a_{w+}=a_{w-}^*$ and $b_{w+}=b_{w-}^*$. The wall channels have the same Wronskian, $2\ii\wtom$, as $u_\pm$, so their coefficients satisfy \eqref{eq:det}, $\operatorname{Im}(b_{w-}a_{w-}^*)=\alpha/(4h-2)$, and the imaginary part of $G_q$ for the wall channel gives
\begin{equation}
\begin{aligned}
\rho_q(\omega,k)&=\frac{(1-K_w^2)\alpha}
{(2h-1)|a_{w-}+K_wa_{w+}|^2}\\
&=\frac{1-K_w^2}{|1-K_w\mathcal R_w|^2}
\rho_{\mathrm{abs},w}(\omega,k),
\end{aligned}
\label{eq:spectral}
\end{equation}
where $\rho_{\mathrm{abs},w}=2\operatorname{Im}G_{\mathrm{abs},w}=\alpha/[(2h-1)|a_{w-}|^2]$ is the finite-wall absorber spectrum, which differs from the thermal spectrum $\rho_0=2\operatorname{Im}G_0\propto\sinh(\pi\alpha)|\Gamma(h+\ii\alpha_L)\Gamma(h+\ii\alpha_R)|^2$ \cite{Son:2002sd,Birmingham:2001pj}. Both carry the sign of $\wtom$, but their magnitudes coincide in the near-horizon limit.

As the AdS boundary is lossless, a wave leaving the wall returns with the same amplitude, $|\mathcal R_w|=1$ on the real axis. At fixed $k$ and wall radius, we write $\mathcal R_w=e^{\ii\Phi_w(\omega)}$, with a continuous real loop phase $\Phi_w$. Its frequency derivative is the Wigner-Smith delay \cite{Wigner:1955zz,Smith:1960zza}
\begin{equation}
\tau_w(\Omega)=
\left.\partial_\omega\Phi_w(\omega)\right|_{\omega=\Omega}>0,
\qquad \Omega-m\Omega_H\neq0,
\label{eq:tau}
\end{equation}
which is the round-trip time of a narrow-band packet centered at a real frequency $\Omega$. Positivity is a consequence of the stored energy (End Matter), as in Foster's reactance theorem \cite{Foster:1924}. The stored energy also provides the basis for the dwell-time relation \cite{Winful:2003zz}. Two geometric times serve for comparison, namely $\tau_0=-2x_w$, the round trip between the wall and the AdS boundary $x=0$ if $x$, a tortoise coordinate near the horizon, were one everywhere along the path, and the round trip $\tau_{\rm null}=2\int_{r_w}^\infty dr/N^2$ of a null ray with zero angular momentum, which exceeds $\tau_0$ by a constant difference. The exact delay is dispersive, with $\tau_w=\tau+O(z_w|\ln z_w|)$ at regular fixed frequencies, not uniformly near corotation (End Matter), where $\tau=\tau_0+\partial_\omega\arg\mathcal R$ has the closed form \eqref{em:targetdelay} and $\mathcal R=-a_+/a_-$ is the exterior reflection normalized at the horizon. At high frequency $\tau_w$ oscillates about $\tau_{\rm null}$ with relative amplitude $r_w/r_+-1=O(z_w)$. Placing the wall close to the horizon does not eliminate exterior dispersion.

\begin{figure}[t]
\includegraphics[width=\columnwidth]{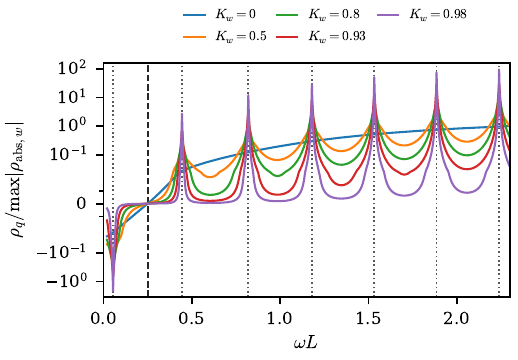}
\caption{Exact spectrum of the local passive wall. The reduced spectral function $\rho_q=2\operatorname{Im}G_q$ is normalized by the largest magnitude of the finite-wall absorber spectrum $\rho_{\mathrm{abs},w}$ (the $K_w=0$ curve) on the plotted range, reached at $\omega L=2.30$; the axis is symmetric-logarithmic, linear below $0.025$. For $K_w>0$, $\rho_q$ is $\rho_{\mathrm{abs},w}$ times the positive Airy filter $\mathcal P_{K_w}$. Dotted lines indicate the Neumann cavity frequencies; the dashed line marks corotation. The mode at $\Omega L\simeq0.054$ lies in the superradiant band and is inverted. Parameters are $L=r_+=1$, $r_-=0.25$, $m=1$, $h=0.75$, $x_w=-8$, and $q=(1-K_w)/(1+K_w)$.}
\label{fig:response}
\end{figure}

Equation \eqref{eq:spectral} has the Fabry-P\'erot form, with the positive filter function $\mathcal P_{K_w}=(1-K_w^2)/(1-2K_w\cos\Phi_w+K_w^2)$. The filter amplifies peaks by a factor $(1+K_w)/(1-K_w)$ and reduces troughs by the inverse factor. At any fixed nonresonant frequency, $\rho_q\to0$ as $K_w\to1^-$. The cavity comb instead arises as a limit of spectral measures 
\begin{equation}
\begin{aligned}
\mathcal P_{K_w}(\Phi_w)&\xrightarrow[K_w\to1^-]{}
2\pi\sum_\nu\delta(\Phi_w-2\pi\nu),\\
\rho_q(\omega)&\xrightarrow[K_w\to1^-]{}
2\pi\sum_\nu\frac{\rho_{\mathrm{abs},w}(\Omega_\nu)}
{|\tau_w(\Omega_\nu)|}\delta(\omega-\Omega_\nu),
\end{aligned}
\label{eq:comb}
\end{equation}
where $\Phi_w(\Omega_\nu)=2\pi\nu$ defines the simple cavity roots. Increasing $K_w$ narrows the filter peaks while without changing their integrated phase weight. Multiplying by $\rho_{\mathrm{abs},w}$ gives the signed comb that the $K_w\to1$ curves of Fig. \ref{fig:response} approach.

The nonnegative filter preserves the sign of the absorbing spectrum, so $\rho_q$ carries the sign of $\wtom$ for every $m$. In the superradiant band $\omega\wtom<0$, which is $0<\omega<m\Omega_H$ for $m\Omega_H>0$ and $m\Omega_H<\omega<0$ for $m<0$ \cite{Zeldovich:1971ffh,Press:1972zz,Brito:2015oca}, the spectral function thus has the sign opposite to $\omega$ for $K_w<1$, and the cavity modes appear as inverted resonances. At $K_w=1$ this holds for the comb measure. The relation $\wtom\rho_q\geq0$ encodes nonnegative dissipation in the corotating frame. On its own, negative spectral weight demonstrates neither scattering amplification nor modal instability.

For the corotating wall, $\chi$ is timelike on $r\geq r_w$, and a positive definite twisted energy [End Matter, \eqref{em:spatialform}--\eqref{em:flux}], valid for every $h>1/2$ including the BF window, is nonincreasing when the AdS source vanishes. We thereby rule out growing scalar modes under standard AdS quantization \cite{Hawking:1999dp,Warnick:2012fi,Holzegel:2012wt} in the End Matter, and for $q>0$ every nonzero normalizable mode has $\operatorname{Im}\omega<0$. A wall rotating at $\Omega_w\neq\Omega_H$ is not covered by this energy argument. This is distinct from the ergoregion instability of horizonless compact objects, which absorption can eliminate \cite{Maggio:2017ivp,Maggio:2018ivz}, from spectral sensitivity to perturbations of the wall \cite{Jaramillo:2020tuu,Cheung:2021bol}, and from Dirichlet \cite{Ortiz:2011wd} and Robin \cite{Dappiaggi:2017pbe,Long:2026vpw,Wang:2026esp} conditions at AdS infinity.

A linewidth alone conflates the loss per return with the propagation delay. With $\lambda=-\ln K_w$ the logarithmic amplitude loss per return and $\mathcal R_w=e^{\ii\Phi_w}$ continued off the real axis, \eqref{eq:poleeq} reads $\Phi_w(\omega_\nu)=2\pi\nu-\ii\lambda$, so that at weak loss the amplitude damping rate $\gamma_\nu=-\operatorname{Im}\omega_\nu$ of the pole near the cavity root $\Omega_\nu$ obeys $\gamma_\nu\tau_w(\Omega_\nu)=\lambda+O(\lambda^3)$. This is the well-known cavity relation \cite{Mark:2017dnq,Macedo:2018yoi,Xie:2026yjg}, in which the damping rate is the loss per return divided by the round-trip time, here the exact dispersive delay. If the delay varies slowly between neighboring cavity modes with spacing $\Delta\Omega_\nu$, then $\gamma_\nu/\Delta\Omega_\nu\simeq\lambda/(2\pi)$ \cite{Macedo:2018yoi}.

\textcolor{mBlue}{\textit{Identifiability without the wall position---}}
Suppose $x_w$ is not specified. In the phase coordinate the radial equation is $\partial_x^2R+\Lambda R=0$ with $\Lambda=[r^2(\omega+mN^\phi)^2-N^2(m^2+\mu^2r^2)]/r_+^2\to\wtom^2$ at the horizon. At any candidate radius $x_t$ with $\wtom U(x_t)\neq0$, the law $q_t(\omega)=\partial_xU/(-\ii\wtom U)|_{x_t}$ reproduces $G_q$ exactly, since $U$ obeys $B_{q_t}U=0$ there. At real frequencies the flux $\operatorname{Im}(\bar U\partial_xU)$ is independent of $x$ and equals $-q\wtom|U(x_w)|^2$ at the wall, so that
\begin{equation}
\operatorname{Re}q_t=q\,\frac{|U(x_w)|^2}{|U(x_t)|^2},\qquad
\operatorname{Im}q_t=\left.\frac{\partial_x\ln|U|^2}{2\wtom}\right|_{x_t},
\label{eq:trial}
\end{equation}
and near the horizon $q_t=(q-\ii\tan\wtom\delta)/(1-\ii q\tan\wtom\delta)+O(\max(z_t,z_w))$ with $\delta=x_t-x_w$. This is the impedance seen through a transmission line of electrical length $\wtom\delta$. Without a restriction of the impedance class the position is therefore not identifiable. Passivity constraints the class. By \eqref{eq:trial} a real law requires a stationary intensity at the wall at every frequency, which a displaced radius satisfies only where $\sin2\wtom\delta=0$, and a constant law requires $\delta=0$ or $K_w=0$. Beyond leading order (End Matter), for a known exterior and fixed $m$, two walls of constant impedance $q\in[0,\infty]$ whose responses coincide on a band of real frequencies have the same position and impedance. The proof uses $\wtom=\ii s$, $s\to\infty$, where every wall is seen from $x_t$ as the local absorber, $\operatorname{Re}q_t\to r_t/r_+-N^2(r_t)/(2r_+s)$; the second term, the nonuniformity of the line, by itself uniquely determines the position of a matched wall.

Conditioning follows from the exact first-order response to the wall parameters,
\begin{equation}
\partial_qG_q=-\frac{\ii\wtom U_w^2}{\mathscr W},\qquad
\partial_{x_w}G_q=\frac{(\Lambda_w-q^2\wtom^2)U_w^2}{\mathscr W},
\label{eq:jacobian}
\end{equation}
where $\mathscr W=v_{1-h}\partial_xv_h-v_h\partial_xv_{1-h}=-2\kappa(2h-1)$ is the Wronskian of the boundary solutions, $U_w=U(x_w)=\mathscr W/B_qv_h$ by \eqref{eq:exactG}, and $|U_w|^2=\kappa(2h-1)\rho_q/(q\wtom)$ by flux conservation. So $|\partial_{\ln q}G_q|=|\rho_q|/2$ and $|\partial_{x_w}G_q|=\Xi|\partial_{\ln q}G_q|$ with $\Xi=|\Lambda_w-q^2\wtom^2|/(q|\wtom|)$, and at every real frequency the two directions are orthogonal in the complex $G_q$ plane. For $N$ samples of $G_q$ with independent errors of rms $\sigma$, the Fisher matrix in $(\ln q,x_w)$ is therefore diagonal, with Cram\'er-Rao bounds
\begin{equation}
\delta\ln q=\frac{\sqrt2\,\sigma}{\|\rho_q\|},\qquad
\delta x_w=\frac{\sqrt2\,\sigma}{\|\Xi\rho_q\|},\qquad
\|f\|^2=\sum_jf(\omega_j)^2.
\label{eq:cr}
\end{equation}
Loss and position are read without trade-off, with weight $\rho_q^2$ concentrated at the Airy peaks. At corotation $\rho_q$ vanishes linearly, so the loss information vanishes as $\wtom^2$, while the position sensitivity retains $O(z_w)$ lower limit determined by the static potential. Near matching with $K_w\to0$, $\Xi=(1-q^2)|\wtom|/q+O(z_w)$ with $\Lambda_w-\wtom^2=\kappa L^2z_w(\omega^2-k^2-\kappa r_+\mu^2)/r_++O(z_w^2)$, so a matched wall is located only at order $z_w=e^{2\kappa x_w}$.

\begin{figure}[t]
\includegraphics[width=\columnwidth]{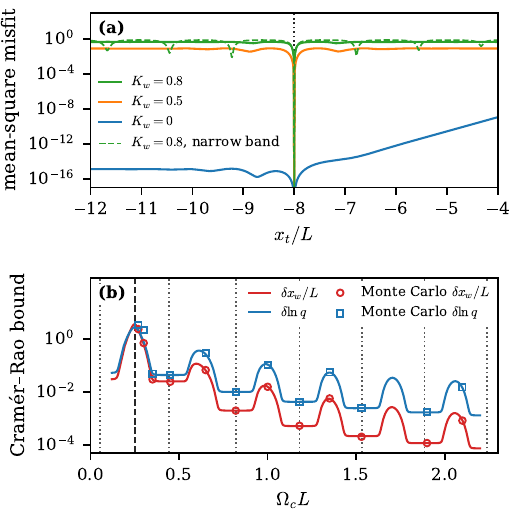}
\caption{Identifiability from a band-limited response on the background of Fig. \ref{fig:response}. (a) Mean-square misfit between the response of the wall at $x_w=-8L$ and that of a wall at $x_t$ with its best constant impedance, in units of $\max\rho_{\mathrm{abs},w}^2$, for the band of Fig. \ref{fig:response} ($200$ frequencies) and for $|\omega-1.53/L|\leq0.1/L$ (dashed). (b) Bounds \eqref{eq:cr} for sliding bands of width $0.2/L$ centered at $\Omega_c$, with $N=64$, $\sigma=0.02\max|\rho_{\mathrm{abs},w}|$, and $K_w=0.8$, against the standard deviations of Monte Carlo least-squares fits to $1000$ noisy realizations (symbols); guides as in Fig. \ref{fig:response}.}
\label{fig:ident}
\end{figure}

Figure \ref{fig:ident}(a) displays the misfit of the best constant-impedance wall at each trial radius, with a unique global minimum at $x_w$ for $K_w>0$, local minima of a narrow band at the fringes $x_w+n\pi/(2\wtom)$ (odd $n$ with the quarter-wave inverted impedance $1/q$), and a landscape flat to $O(z^2)$ for $K_w=0$. Figure \ref{fig:ident}(b) tests \eqref{eq:cr} for bands centered at $\Omega_c$ by simulated measurements, in which complex Gaussian noise of rms $\sigma$ is added to the exact $G_q$ at the sampled frequencies and $(\ln q,x_w)$ are fitted by Monte Carlo least squares, $1000$ times per band. Where the bounds are below $0.1$ the dispersion of the fits reaches them to within $5\%$ and loss and position are uncorrelated; near corotation the fits are biased. The bounds increase by three to four orders of magnitude from a band on a resonance to one at corotation. The induced ratio supplies the global starting point, $|\zeta_{\rm loc}|=K_w+O(z_w)$ and $\partial_\omega\arg\zeta_{\rm loc}=\tau_0+O(z_w|\ln z_w|)$.

\textcolor{mBlue}{\textit{Outlook---}}
The spectral function of a local passive wall is the finite-wall absorber spectrum multiplied by a positive Airy filter, and a positive energy excludes growing modes and fixes the sign of the return delay. Once the exterior and the boundary conventions are fixed, the complex boundary response determines the induced horizon return amplitude. Within the class of constant impedances, its values on any band also fix the wall position and impedance, which move the response in orthogonal directions in the complex plane. The boundary thus sees a reflecting wall, whereas a perfectly absorbing wall is hidden up to terms exponentially suppressed in the round-trip time. These results provide analytic reference points for cavity and echo calculations \cite{Cardoso:2019rvt,Mark:2017dnq,Chen:2025jbf} and for analogue gravity experiments \cite{Torres:2020tzs,Svancara:2023yrf} and their modeling \cite{Patrick:2025tpk}. For a frequency-dependent law shared by all azimuthal numbers, rotation determines the position only up to displacements $n\pi/\Omega_H$ at leading order near the horizon; this class, the boundary noise of the closed realization in the thermal (Kubo-Martin-Schwinger) state of the corotating frame, and a causality theorem for the exact response remain open.

\textit{Acknowledgments---}
We thank Sizheng Ma for reading the manuscript and for helpful discussions. This work was supported by the National Natural Science Foundation of China (Grant No. 12365010) and the Natural Sciences and Engineering Research Council of Canada (NSERC). J.Y. acknowledges support from NSERC through a Canada Graduate Research Scholarship – Doctoral (Funding Reference No.\ 110\_2026\_2027\_Q1\_571). Z.Z. acknowledges financial support from the MUR FIS2 Advanced Grant ET-NOW (CUP:~B53C25001080001) and the INFN TEONGRAV initiative. Research at Perimeter Institute is supported in part by the Government of Canada through the Department of Innovation, Science, and Economic Development and by the Province of Ontario through the Ministry of Colleges and Universities. 

\textit{Note added---}
While this work was in the final stage of preparation, Ref. \cite{Long:2026ngh} appeared, which has some overlap with our settings.

%

\clearpage
\setcounter{equation}{0}
\renewcommand{\theequation}{A\arabic{equation}}
\renewcommand{\theHequation}{endmatter.\arabic{equation}}

\section*{End Matter}
\label{sec:endmatter}

\subsection*{Source convention and wall channels}
For the scalar action
\begin{equation}
S_\varphi=-\frac{C_\varphi}{2}\int d^3x\sqrt{-g}
(g^{ab}\partial_a\varphi\partial_b\varphi+\mu^2\varphi^2),
\label{em:action}
\end{equation}
$C_\varphi>0$ sets the operator normalization and $g$ is the metric determinant. Writing the asymptotic source and response coefficients as $\varphi_s,\varphi_v$ in $\varphi=\varphi_s r^{2h-2}+\varphi_v r^{-2h}+\cdots$, the renormalized ultraviolet boundary contribution is $\delta S_{\rm ren}=C_\varphi(4h-2)L^{-2}\int dt\,d\phi\,\varphi_v\delta\varphi_s$, up to local terms \cite{Skenderis:2002wp}. With the operator defined using this measure, the Hamiltonian coupling $+\mathcal J\mathcal O$ implies $\varphi_s=-\mathcal J$. Setting $D=r_+^2-r_-^2$, the boundary coefficients $a,b$ satisfy $\varphi_s=D^{1-h}a$, $\varphi_v=D^hb$. Hence $\mathcal N_h=C_\varphi(4h-2)D^{2h-1}/L^2>0$ and $G_R=-\mathcal N_hG_q+P$, with $-2\operatorname{Im}G_R=\mathcal N_h\rho_q$ for real $P$.

The on-shell variation of \eqref{em:action} also yields an inner-wall term $C_\varphi r_+\int dt\,d\vartheta\,\partial_x\varphi\,\delta\varphi|_{r_w}$, $\vartheta=\phi-\Omega_Ht$, and no ordinary real single field boundary action yields the dissipative wall law. A closed realization attaches at each corotating angle a semi-infinite channel $\psi(t,y,\vartheta)$, $y<0$, with action $\tfrac12C_\varphi\int Z_c[(\chi\psi)^2-(\partial_y\psi)^2]$, $\psi(t,0,\vartheta)=\varphi(t,r_w,\vartheta)$, and an outgoing condition as $y\to-\infty$. Stationarity cancels the inner-wall term and gives $r_+\partial_x\varphi=Z_c\chi\varphi$ at $r_w$, i.e., $q=Z_c/r_+$. The exterior loses energy as flux into the channel, whose outgoing condition serves as the ingoing horizon condition. An incoming channel wave would give $q=-Z_c/r_+$. The channel state, and with it the noise, is chosen separately.

The boundary falloffs determine the Wronskian with respect to $x$, $\mathscr W=v_{1-h}\partial_xv_h-v_h\partial_xv_{1-h}=-2\kappa(2h-1)$. The wall coefficients are
\begin{equation}
\begin{aligned}
a_{w\pm}&=(\partial_xv_h\mp\ii\wtom v_h)/\mathscr W,\\
b_{w\pm}&=(-\partial_xv_{1-h}\pm\ii\wtom v_{1-h})/\mathscr W,
\end{aligned}
\label{em:wallcoeffs}
\end{equation}
evaluated at $r_w$. Their determinant is $-\ii\alpha/(2h-1)$, and $a_{w-}+K_wa_{w+}=(1+K_w)B_qv_h/\mathscr W$. These equalities give \eqref{eq:factorized} and \eqref{eq:spectral}. In equivalent form, at regular real frequencies,
\begin{equation}
\rho_q=\frac{1-|\zeta_{\rm loc}|^2}{|1-\zeta_{\rm loc}\mathcal R|^2}\rho_0,
\label{em:thermalfilter}
\end{equation}
where flux conservation implies $1-|\zeta_{\rm loc}|^2=q|u_-(r_w)+\zeta_{\rm loc}u_+(r_w)|^2\geq0$ for unit ingoing horizon amplitude and $\wtom\neq0$.

\subsection*{Positive energy and mode stability}
In the corotating coordinates $(t,r,\vartheta=\phi-\Omega_Ht)$, $\chi=\partial_t$. Define the radial weights $p=rN^2$, $\varpi=r/N^2$, and $c_r=1/r-\varpi(\Omega_H+N^\phi)^2>0$, with $N^\phi$ from \eqref{eq:metric}, where the positivity of $c_r$ is a consequence of
\begin{equation}
g(\chi,\chi)=-\frac{(r_+^2-r_-^2)(r^2-r_+^2)}{L^2r_+^2}<0.
\label{em:timelike}
\end{equation}
In what follows, dots and primes represent $\partial_t$ and $\partial_r$, respectively, and a subscript $w$ denotes evaluation at the wall. Since $\partial_x=(p/r_+)\partial_r$, the field equation and the wall condition read
\begin{equation}
\begin{aligned}
\varpi\ddot\varphi-2\varpi(\Omega_H+N^\phi)\partial_\vartheta\dot\varphi
-(p\varphi')'-c_r\partial_\vartheta^2\varphi+r\mu^2\varphi&=0,\\
p_w\varphi'_w=q r_+\dot\varphi_w.&
\end{aligned}
\label{em:evolution}
\end{equation}
To allow for negative $\mu^2$, we choose the static solution $f=y^h{}_2F_1(h,h;2h;y)$, $y=1-z$, satisfying $(pf')'=r\mu^2f$. Since the series coefficients are positive and $y(r)$ is decreasing, $f>0$ and $f'<0$ for $h>1/2$. For standard source-free solutions $u$,
\begin{equation}
\begin{aligned}
\int_{r_w}^{\infty}(p|u'|^2+r\mu^2|u|^2)dr
={}&\int_{r_w}^{\infty}pf^2|(u/f)'|^2dr\\
&-p_w\frac{f'_w}{f_w}|u_w|^2.
\end{aligned}
\label{em:twistidentity}
\end{equation}
The dropped boundary term is $O(r^{2-4h})\to0$. Thus the spatial form
\begin{equation}
\begin{aligned}
\mathfrak a[u]={}&\int_{r_w}^{\infty} dr\,\int_0^{2\pi} d\vartheta
[pf^2|(u/f)'|^2+c_r|\partial_\vartheta u|^2]\\
&-p_w\frac{f'_w}{f_w}\int_0^{2\pi} d\vartheta\,|u_w|^2
\end{aligned}
\label{em:spatialform}
\end{equation}
is positive for $u\neq0$. Define $E=\tfrac12[\int dr\,d\vartheta\,\varpi|\dot\varphi|^2+\mathfrak a[\varphi]]$. Multiplying \eqref{em:evolution} by $\dot\varphi^*$ and integrating gives
\begin{equation}
\begin{aligned}
\dot E&=-qr_+\int d\vartheta\,|\dot\varphi_w|^2\leq0,\\
2\operatorname{Im}\omega E_0&=-2\pi qr_+|\wtom u_w|^2,
\end{aligned}
\label{em:flux}
\end{equation}
where $E_0>0$ is the energy of a nonzero mode $\varphi=e^{-\ii\wtom t+\ii m\vartheta}u(r)$ with its exponential time factor factored out. The gyroscopic term has zero integral and the AdS flux vanishes \cite{Hawking:1999dp,Warnick:2012fi,Holzegel:2012wt}, so no mode grows. For $q>0$, a real mode with $\wtom\neq0$ would have to satisfy $u_w=u'_w=0$ and would therefore vanish everywhere. For $\wtom=0$, the static equation gives $\mathfrak a[ue^{\ii m\vartheta}]=0$. As a result, for $q>0$, every nonzero normalizable mode has $\operatorname{Im}\omega<0$, whereas $q=0$ admits real Neumann modes.

Before imposing the wall condition, we apply the same energy balance to a normalizable exterior mode. With $s=-\ii\wtom$ and $\mathcal E=E_0/(2\pi)$ its energy per unit angle, its wall amplitudes satisfy
\begin{equation}
2\operatorname{Re}s\,\mathcal E =r_+|s|^2(|C_{\rm out}|^2-|C_{\rm in}|^2).
\label{em:exteriorflux}
\end{equation}
For $\operatorname{Re}s>0$, positivity and analyticity of $v_h$ give $C_{\rm out}\neq0$ and $|\mathcal R_w|=|C_{\rm in}/C_{\rm out}|<1$, which precludes zeros of $1-K_w\mathcal R_w$ in the upper half plane. At a regular real frequency, normalize $C_{\rm out}=1$, then $|\mathcal R_w(\omega+\ii\epsilon)|^2=1-2\epsilon\tau_w+O(\epsilon^2)$ gives $\mathcal E=r_+\wtom^2\tau_w>0$ for $\wtom\neq0$. 

\subsection*{Residues and the near-horizon limit}
At a simple physical pole,
\begin{equation}
\operatorname*{Res}_{\omega_\nu}G_q =-\left.\frac{B_qv_{1-h}}{\partial_\omega(B_qv_h)}\right |_{\omega_\nu},
\qquad \partial_\omega(B_qv_h)\neq0.
\label{em:targetresidue}
\end{equation}
The derivative accounts for the frequency dependence of $B_q$ itself. A nonzero Wronskian $\mathscr W$ excludes shared zeros of $B_qv_h$ and $B_qv_{1-h}$. The regular $v_\beta$ basis also remains valid at $\wtom=0$. Degeneracies of the horizon basis cancel through the identity
\begin{equation}
a_-B_qu_+-a_+B_qu_-=-\frac{\ii\alpha}{2h-1}B_qv_h.
\label{em:cancellation}
\end{equation}
The approximate constant horizon ratio need not retain this cancellation. In a nonextremal background with fixed $x_w<0$ and $0<K_w\leq1$, the constant-modulus ratio $\zeta_{\rm app}=K_we^{-2\ii\wtom x_w}$ gives rise to poles near $\alpha=\ii n$, $n\geq1$. There $a_+/a_-$ has a simple pole from $\Gamma(1+\ii\alpha)$, with residue $A_n$, while $\zeta_{\rm app}=K_wz_w^n\neq0$, so $1+\zeta_{\rm app}a_+/a_-$ vanishes at $\alpha_*=\ii n+\eta$ with $\eta\simeq-\zeta_{\rm app}A_n$, where the numerator equals $-\ii\alpha_*\zeta_{\rm app}/[(2h-1)a_-(\alpha_*)]\neq0$ by \eqref{eq:det}. For $m\neq0$ this holds at every sufficiently large $n$ while for $m=0$ and $r_->0$ the pole of $a_+$ is cancelled at the exceptional $n$ with $h-(r_+\pm r_-)n/(2r_+)\in\mathbb Z_{\leq0}$.

For fixed regular real frequencies and fixed $K_w<1$,
\begin{equation}
\begin{aligned}
\zeta_{\rm loc}&=K_we^{-2\ii\wtom x_w}+O(z_w),\\
\mathcal R_w&=\mathcal R e^{\ii\wtom\tau_0}+O(z_w),\quad
G_{\mathrm{abs},w}=G_0+O(z_w).
\end{aligned}
\label{em:nearhorizon}
\end{equation}
These error terms need not be uniform near degeneracies or corotation, nor jointly with the lossless resonance limit. Differentiating the wall phase $e^{-2\ii\wtom x_w}$ in \eqref{em:nearhorizon} with respect to $\omega$ introduces a factor of $x_w=\ln z_w/(2\kappa)$, so $\tau_w=\tau+O(z_w|\ln z_w|)$. Differentiating the horizon coefficients gives the comparison delay
\begin{equation}
\tau=\tau_0+\frac2\kappa\operatorname{Re}\psi(1+\ii\alpha)
-\sum_{a=L,R}\frac{\operatorname{Re}\psi(h+\ii\alpha_a)}{2\pi T_a},
\label{em:targetdelay}
\end{equation}
where $\psi$ is the digamma function. For $\omega\to\infty$, $\operatorname{Re}\psi(1+\ii\alpha)\to\ln\alpha$ and $\operatorname{Re}\psi(h+\ii\alpha_a)\to\ln\alpha_a$, so $\tau\to\tau_0+\sum_a(2\pi T_a)^{-1}\ln(4\pi T_a/\kappa)=\tau_{\rm null}+O(z_w)$, using $(2\pi T_L)^{-1}+(2\pi T_R)^{-1}=2/\kappa$.

On the real axis the wall determinant gives $G_{\mathrm{abs},w}(1+V_w/\mathcal R_w)=\ii\alpha/[(2h-1)a_{w-}a_{w+}]=\ii\rho_{\mathrm{abs},w}$, regular away from corotation because a real zero of $a_{w-}\propto B_1v_h$ would be a real mode of the $q=1$ wall, which the energy identity rules out.

\subsection*{Identifiability and conditioning}

Let walls $(x_1,q_1)$ and $(x_2,q_2)$ with constant $q_j\in[0,\infty]$ produce responses that conincide on a set of real frequencies with an accumulation point. Being meromorphic, the response functions agree identically, and so do the unit-source solutions $U$. If $x_1=x_2$, the wall conditions give $(q_1-q_2)\wtom U(x_1)=0$ with $U(x_1)=\mathscr W/B_{q_1}v_h\neq0$ away from poles, so $q_1=q_2$. If $x_1<x_2$, then $Y=\partial_xU/U$ satisfies the Riccati equation $Y'=-\Lambda-Y^2$ on $[x_1,x_2]$ with $Y(x_1)=-\ii q_1\wtom$ and $Y(x_2)=-\ii q_2\wtom$ for every $\omega$. Put $\wtom=\ii s$, $s\to+\infty$, so that $-\Lambda=s^2g_1-\ii sg_2+g_3$ with $g_1=r^2/r_+^2$, $g_2=2mr^2(\Omega_H+N^\phi)/r_+^2$, and $g_3=[N^2(m^2+\mu^2r^2)-m^2r^2(\Omega_H+N^\phi)^2]/r_+^2$. The half plane $\operatorname{Re}Y\geq0$ is forward invariant, since $\operatorname{Re}Y'=s^2g_1+g_3-(\operatorname{Re}Y)^2+(\operatorname{Im}Y)^2>0$ on its boundary, and contains $Y(x_1)=q_1s$ (from $+\infty$ for $q_1=\infty$). The Liouville-Green branch $Y_+=s\sqrt{g_1}-g_1'/(4g_1)-\ii g_2/(2\sqrt{g_1})+O(1/s)$ solves the same equation; $\delta=Y-Y_+$ satisfies $\delta'=-(Y+Y_+)\delta$ with $\operatorname{Re}(Y+Y_+)\geq s\sqrt{g_1}-O(1)$, so $Y(x_2)=Y_+(x_2)+O(se^{-cs})$, $c>0$. Since $g_1'=2r^2N^2/r_+^3$, $\operatorname{Re}Y(x_2)=sr_2/r_+-N^2(r_2)/(2r_+)+O(s^{-1})$, which cannot equal $q_2s$ for all $s$ because $N^2(r_2)>0$; $q_2=\infty$ require $U(x_2;\omega)\equiv0$.

Equation \eqref{eq:jacobian} is the first-order response $\delta G_q=-U_w(\delta BU)_w/\mathscr W$ to a variation $\delta B$ of $B_q$, from \eqref{eq:exactG} and $U_w=\mathscr W/B_qv_h$. Here $\delta B=\ii\wtom\,\delta q$ for the impedance, and $\delta B=\delta x_w\,\partial_xB_q$ with $\partial_xB_qU=-(\Lambda_w-q^2\wtom^2)U$ on the wall for a position shift. The intensity relation is the Wronskian $U\partial_x\bar U-\bar U\partial_xU=-2\ii\mathscr W\operatorname{Im}G_q=2\ii q\wtom|U_w|^2$ at real $\omega$ ($v_\beta$ real). For independent circular complex Gaussian errors, $\langle|\delta G_q|^2\rangle=\sigma^2$, the Fisher matrix $\mathcal I_{ab}=(2/\sigma^2)\sum_j\operatorname{Re}(\overline{\partial_aG_q}\,\partial_bG_q)$ is diagonal, because $\overline{\partial_{\ln q}G_q}\,\partial_{x_w}G_q\propto\ii q\wtom(\Lambda_w-q^2\wtom^2)|U_w|^4$ is imaginary.

\end{document}